\documentstyle[epsfig]{mn}
\begin{document}

\def\ltsima{$\; \buildrel < \over \sim \;$}
\def\simlt{\lower.5ex\hbox{\ltsima}}
\def\gtsima{$\; \buildrel > \over \sim \;$}
\def\simgt{\lower.5ex\hbox{\gtsima}}
\def\ls{{_<\atop^{\sim}}}
\def\lax{{_<\atop^{\sim}}}
\def\gs{{_>\atop^{\sim}}}
\def\gax{{_>\atop^{\sim}}}
\def\cgs{ ${\rm erg~cm}^{-2}~{\rm s}^{-1}$ } 
 
\title[The HELLAS survey]
{The BeppoSAX High Energy Large Area Survey HELLAS, II: 
Number counts and X-ray spectral properties}

\author[F. Fiore et al.]{F. Fiore$^{1,2,3}$, P. Giommi$^1$, 
C. Vignali$^{4,5}$, A. Comastri$^5$, G. Matt$^6$, G.C. Perola$^6$,
\\ ~ \\
{\LARGE F. La Franca$^6$, S. Molendi$^7$, F. Tamburelli$^1$,
and L.A. Antonelli$^2$}\\ ~ \\
$^1$ BeppoSAX Science Data Center, Via Corcolle 19, I--00131 Roma, Italy\\
$^2$ Osservatorio Astronomico di Roma, Via Frascati 33,
I--00044 Monteporzio, Italy\\
$^3$ Harvard-Smithsonian Center of Astrophysics, 60 Garden Street, 
Cambridge MA 02138 USA\\
$^4$ Dipartimento di Astronomia, Universit\`a di Bologna, via Ranzani 1, 
I--40127 Bologna, Italy \\
$^5$ Osservatorio Astronomico di Bologna, via Ranzani 1, I--40127
Bologna, Italy \\
$^6$ Dipartimento di Fisica, Universit\`a degli Studi ``Roma Tre",
Via della Vasca Navale 84, I--00146 Roma, Italy \\
$^7$ IFCTR/CNR, via Bassini 15, Milano, I--20133, Italy\\
}

\maketitle
\begin{abstract}
The BeppoSAX High Energy Large Area Survey (HELLAS) has surveyed about
85 deg$^2$ of sky in the 5--10 keV band down to a flux of
$4-5\times10^{-14}$ \cgs.  The source surface density of 16.9$\pm$6.4
deg$^{-2}$ at the survey limit corresponds to a resolved fraction of
the 5--10 keV X--ray background (XRB) of the order of 20--30 \%.

Hardness ratios analysis indicates that the spectra of a substantial
fraction of the HELLAS sources (at least one third) are harder than a
$\alpha_E=0.6$ power law. This hardness may be due to large absorbing
columns. The hardness ratio analysis also indicates that many HELLAS
sources may have a spectrum more complex than a single absorbed power
law. A soft component, superimposed to a strongly cut-off power law,
is likely to be present in several sources.

\end{abstract}
 
 
\section{Introduction}

Hard X-ray observations are very efficient in tracing emission due to
accretion mechanisms, like in Active Galactic Nuclei (AGN).  Hard
X-ray selection is not affected by strong biases present at other
wavelengths.  For example, a column of a few times $10^{22}$ cm$^{-2}$
has negligible effect in the 5-10 keV band, while it reduces by
$\sim100$ times nuclear emission below 2 keV. Optical and UV color
selection is biased against objects with even modest extinction or an
intrinsically `red' emission spectrum (see e.g. Vignali et al. 2000,
Maiolino et al. 2000). However, the space density and evolutionary
properties of the hard X-ray sources are still basically unknown.  We
therefore decided to take advantage of the large field of view and
relatively high sensitivity and spatial resolution of the BeppoSAX
MECS (Boella et al. 1997a,b) to perform a large area survey in the
hardest band (5--10 keV) which is reachable so far with imaging
instruments.  Chandra and XMM-Newton go much deeper than BeppoSAX at
low energies, but in the 5-10 keV band the BeppoSAX MECS collecting
area is comparable to that of Chandra ACIS and it is a factor of 5-10
smaller than that of XMM-Newton Epic PN.  However, the MECS field of
view is a factor of about 7 larger than that of the Chandra ACIS-I
instrument and about 3 times larger than that of the XMM Epic PN.
Therefore, these instruments will cover, at least for the first few
years of operations, a relatively small portion of the sky.  The
BeppoSAX shallower but larger area survey nicely complements deep
pencil beam {\it Chandra} and XMM--{\it Newton} surveys (Mushotzky et
al. 2000, Giacconi et al. 2001, Hasinger et al. 2001, Hornshemeier et
al 2001).  The main scientific purpose of the BeppoSAX MECS survey is
therefore to study the nature of the hard X-ray source population and
in particular to provide a relatively large sample of sources bright
enough to measure, with the present generation of X-ray satellites,
their main X-ray spectral properties This will allow for the first
time the statistical study of the distribution and evolution of the
obscuring gas in a hard X-ray selected sample.

Preliminary results on this survey have been presented by Giommi et
al.  (1998), Ricci et al. (1998), Comastri et al. (1999, 2000), Matt
et al. (1998), Fiore et al. (1998, 2000).  These indicates that the
BeppoSAX survey has resolved 20 to 30\% of the hard XRBand that the
large majority of the sources so far identified are AGN in agreement
with optical identifications results of ASCA surveys in the 2--10 keV
energy range (Akiyama et al. 2000).  Interestingly enough the fraction
of obscured AGN among the about 70 optically identified HELLAS sources
is higher than in the ROSAT and ASCA samples (Fiore et al. 1999, Fiore
et al. 2000, La Franca et al. in preparation).

In this paper we present the survey and discuss the sources X--ray
spectral properties.  An analysis of the BeppoSAX MECS position
reconstruction accuracy is presented in the Appendix. A companion
paper (Comastri et al. 2001) compares the observed number counts and
absorbing column density distribution with AGN synthesis models for
the hard X-ray cosmic backgrund (XRB).

\section{The BeppoSAX HELLAS survey}

The High Energy Large Area Survey (HELLAS) has been performed in the
hard 4.5-10 keV, `octave' wide, band for three reasons: a) this is the
band closest to the maximum of the XRB energy density which is
reachable with the current imaging X-ray telescopes; b) the BeppoSAX
MECS (Boella et al. 1997b) Point Spread Function (PSF) greatly
improves with energy: in the 4.5-10 keV band it is a factor of $\sim2$
sharper than in the softer 1.5-4.5 keV band (providing a 95 \% error
radius of 1', see the Appendix and Ricci et al 1998), which allows
optical identification of the sources; c) the MECS background
(on-axis, internal + 70\% Cosmic X-ray background) is $\sim
4.3\times10^{-3}$ counts s$^{-1}$ arcmin$^{-2}$ (3 MECS units) in the
4.5-10 keV band and $\sim 3.0\times10^{-3}$ counts s$^{-1}$
arcmin$^{-2}$ in the 1.3-4.5 keV band. So using the higher energy band
only, reduces the total background by $\approx 40 \%$, enhancing the
chance of detecting faint, highly obscured or very hard sources, with
few counts below 4.5 keV.

About 85 square degrees of sky have been surveyed in the 4.5-10 keV
band using 142 BeppoSAX MECS high Galactic latitude ($|b|>20$ deg)
fields. There is no overlap among the fields and, wherever possible,
multiple observations of the same field have been merged in one single
pointing to increse the sensitivity.  Fields were selected among
public data (as March 1999) and our proprietary data.  Fields centered
on bright extended sources and bright Galactic sources were excluded
from the survey, as well as fields close to LMC, SMC and M33.  Most of
the fields have exposures between 30 and 100 ks, and 20 fields have an
exposure higher than 80 ks (Fig.~1).

\begin{figure}
\centerline{
\epsfig{file=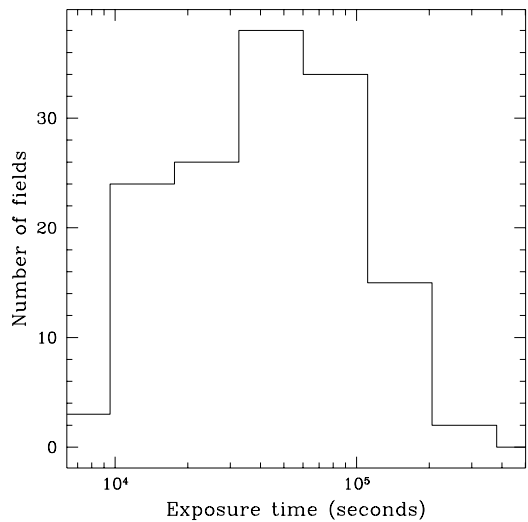, width=8cm, height=8cm}
}
\caption{
The distribution of the net exposure times in the 142 HELLAS fields.
}
\label{expo}
\end{figure}

\subsection {Source detection}

Coadded, sky coordinates MECS1+MECS2+MECS3 images (or MECS2+MECS3
images after the loss of MECS1 on May 7th 1997) have been analysed.
Sources were detected in images accumulated between 4.5 and 10 keV.
The lower limit is chosen to include in the band the Xenon L edges
(the MECS Energy-PI relationship is non-linear accross the
edge). Cleaned and linearized MECS event files in the BeppoSAX SDC on
line archive have always been used in this analysis.  The reader is
referred to the SDC on-line documentation and to Fiore, Guainazzi \&
Grandi (1999) for details on data reduction.

We have used a variation of the DETECT routine included in the XIMAGE
package (Giommi et al. 1991) to detect the X-ray sources.  The method
consists in first convolving the X-ray image with a wavelet function,
to smooth the image and increase contrast, and then in running a
standard slide-cell detection method on the smoothed image, to locate
count excesses above the local background.  The detection algorithm
has been run several times for each field, changing the size of the
slide-cell and the width of the wavelet function to take into account
the variation of the MECS PSF with the off-axis angle, and to improve
the efficiency in detecting extended sources.  The quality of the
detection has always been checked interactively.  In particular,
sources near the MECS Fe$^{55}$ calibration sources have been
carefully tested for reliability running the detection algorithm after
excluding a narrow energy interval centered on the iron feature at
5.894 keV.  The final net counts are estimated from the original
(un-smoothed) image, to preserve Poisson statistics.  The background
is calculated using ten source-free boxes near the source region and
is rescaled at the source position to take into account the spatial
variations of the MECS background (see the BeppoSAX SDC on line
``cookbook'', http://www.asdc.asi.it/bepposax/software/cookbook, and
Chiappetti et al.  1998).  175 sources have been detected in the
4.5-10 keV images with a probability higher than 99.94 \% that they
are not Poisson fluctuation of the background (after excluding the
targets of the observations).  The analysis of the 2-10 keV images is
reported by Giommi, Perri \& Fiore (2000).

\subsection{Count rates and fluxes}

\begin{figure}
\centerline{
\psfig{file=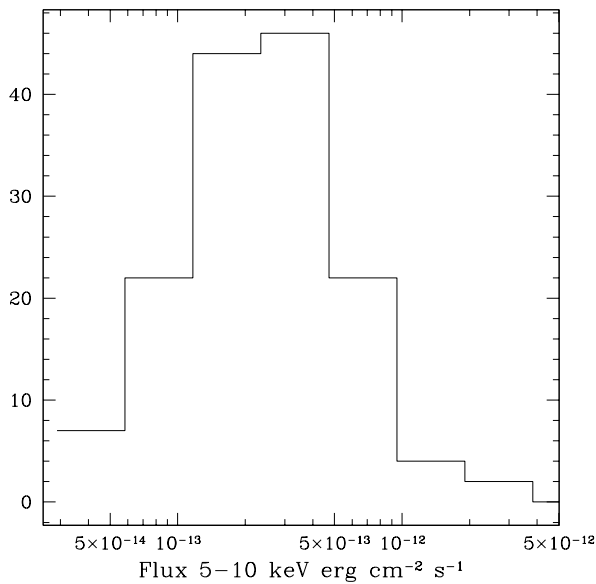, width=9cm, height=9cm}
}
\caption{
The histogram of the number of sources (non-target)
detected as a function of the 5-10 keV flux.
}
\label{fxhisto}
\end{figure}

Source count rates in four bands (1.3--10 keV, total, T, 1.3--2.5 keV,
low band, L, 2.5--4.5 keV, middle band, M, and and 4.5--10 keV, high
band, H) were extracted and corrected for the energy dependent
vignetting (as calibrated in orbit using a series of observations of
the Crab nebula at different off-axis angles, Conti et al. 1997,
Cusumano \& Mineo 1998) and for the MECS PSF (using an analytical
approximation of the PSF calibrated using observations of AGNs and
Compact Galactic sources, Conti et al. 1997).  The MECS 4.5-10 keV PSF
half power radius is $\sim1.2$ arcmin at off-axis angles $\sim7$ arcmin, 
it increases to
$\sim1.65$ arcmin and $\sim2.2$ arcmin at off-axis angles of 18.8
and 24.5 arcmin respectively. Even at these large off-axis angles the
core can be safely considered axisymmetric.  For faint sources
($S/N<4$) all corrections were done at fixed energies, close to the
mean energy for a power law spectrum with $\alpha_E=0.6$ in the three
bands L, M and H. For brighter sources the correction was done by
convolving the vignetting and PSF functions with the source count
histogram in the three L, M and H bands.  The count rates were
converted to fluxes using a conversion factor of 7.8$\times10^{-11}$
\cgs (5-10 keV flux) per one ``3 MECS count'' (4.5-10 keV),
appropriate for a power law spectrum with $\alpha_E=0.6$. The factor
is not strongly sensitive to the spectral shape, due to the narrow
band: for $\alpha_E=0.4$ and 0.8 it is 8.1 and 7.6 $\times10^{-11}$
\cgs, respectively.  A conversion factor of 9.9$\times10^{-11}$ \cgs
per one ``3 MECS count'' has been used for sources under the 550
micron Berillium strongback supporting the MECS window to account for
the reduced detector sensitivity.  The histogram of the number of
sources as a function of the flux is given in figure \ref{fxhisto}.

\subsection{Sky coverage}

\begin{figure*}
\centerline{
\hbox{
\psfig{file=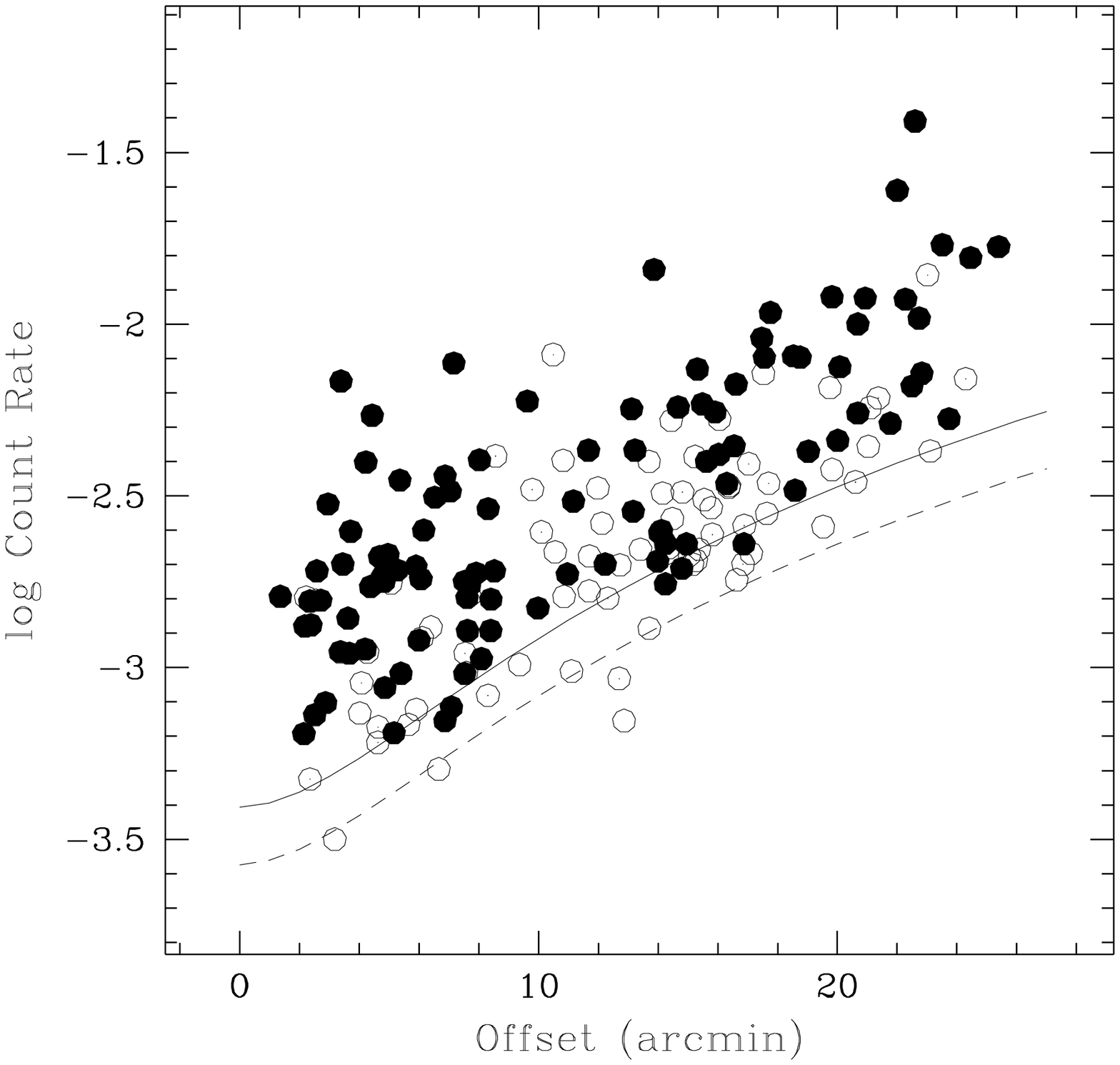, width=9cm, height=9cm}
\psfig{file=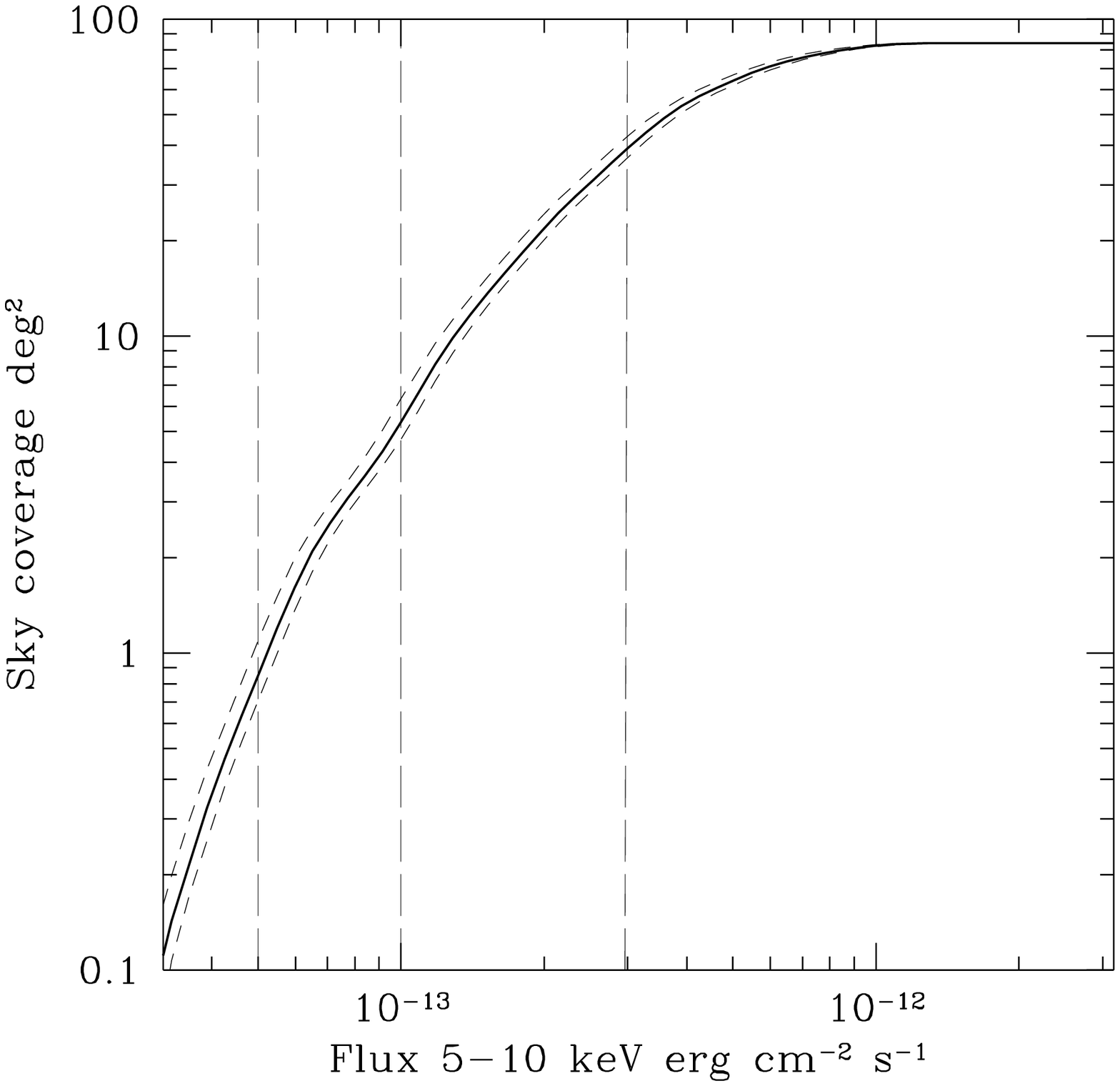, width=9cm, height=9cm}
}
}
\caption{
a) The MECS count rate as a function of the off-axis angle.
Solid points are from observations with
3 MECS units, open points from observations with 2 MECS units.  The
solid (dashed) model line represents the minimum count rate detectable
(at the given threshold of 99.94 \%) for an exposure time of 100 ks in
3 (2) MECS units. 
b) The HELLAS sky coverage for three different models
of the source spectrum: a power law with energy index
of 0.6 (thick solid line), 0.2 and 1.0 (lower and upper
dashed lines).
}
\label{sky}
\end{figure*}

Figure \ref{sky}a) gives, for the 175 HELLAS sources, the count rate
as a function of the off-axis angle.  The solid (dashed) model line
represents the minimum detectable count rate (at the given threshold
of 99.94 \%) for an exposure time of 100 ks in 3 (2) MECS units.
These lines define the count rate limit (function of the exposure time
and off axis angle) of the HELLAS survey.  The shape of the model
curves has been adjusted using the latest calibration of the telescope
vignetting, MECS background and Berillum strongback.  Their
normalization has been chosen using extensive simulations.  The survey
has been simulated assuming a given logN-logS, a distribution of
spectral shapes and the positions in detector coordinates of the
HELLAS sources. The sky coverage of the simulated survey has been
computed inverting the model curves.  The normalization of the model
curves (the on-axis flux limits for a 100 ks observation), and
therefore the sky coverage, were then varied until the input logN-logS
was correctly reproduced by the one computed from the simulations.
The adopted value of the on-axis flux limit is $3\times10^{-14}
(100/t(ks))^{0.5})$ \cgs (3 MECS units). It is $5\times10^{-14}
(100/t(ks))^{0.5})$ and $1.5\times10^{-13} (100/t(ks))^{0.5})$ at
off-axis angles of 6 and 15 arcmin, respectively.  Regions of radius
4, 6 or 8 arcmin around bright targets have been excluded from the sky
coverage and sources detected in these regions have been excluded from
the sample.  The exclusion radius has been determined by imposing that
the target count rate per square arcmin at a given off-axis radius is
less than half of the local MECS background.  Sources with a count
rate smaller than the minimum value given by the model curves in
figure \ref{sky}a) normalized to the actual exposure time have also
been excluded from the sample. This brings the total number of sources
in the sample used for computing the logN-logS to {\bf 147}.  The
HELLAS source catalogue including the source identification number,
the J2000.0 coordinates, the 5--10 keV flux and signal to noise and
the count ratios in different energy ranges is reported in Table 2.
The sky coverage computed assuming a power law spectrum with
$\alpha_E=0.6$ is given in figure \ref{sky}b), thick solid line, and
in table 1.

\begin{table}
\label{tablesky}
\caption{\bf HELLAS sky coverage and integral logN-logS }
\begin{tabular}{lcl}
\hline
Flux 5-10 keV          & sky coverage & logN-logS  \\
\cgs                   & deg$^2$       & deg$^{-2}$ \\
\hline
$1.00\times10^{-12}$ & 84.2   & 0.0359$\pm$0.0207 (0.035)\\
$7.38\times10^{-13}$ & 77.1   & 0.0891$\pm$0.0325 (0.070)\\
$5.45\times10^{-13}$ & 67.7   & 0.253$\pm$0.0579  (0.11)\\
$4.02\times10^{-13}$ & 55.2   & 0.447$\pm$0.0808  (0.21)\\
$2.97\times10^{-13}$ & 37.5   & 0.952$\pm$0.133   (0.32)\\
$2.19\times10^{-13}$ & 25.2   & 1.54$\pm$0.19     (0.42)\\
$1.62\times10^{-13}$ & 15.9   & 2.40$\pm$0.28     (0.68)\\
$1.19\times10^{-13}$ &  8.3   & 4.43$\pm$0.51     (1.1)\\
$8.81\times10^{-14}$ &  4.7   & 5.92$\pm$0.75     (2.2)\\
$6.50\times10^{-14}$ &  2.9   & 10.3$\pm$1.4      (2.8)\\
$4.80\times10^{-14}$ &  0.75  & 16.9$\pm$3.0      (6.4)\\
\hline

\end{tabular}
 
First
quoted errors are the $1\sigma$ statistical confidence
interval. Errors in brackets systematic uncertainties, see the text

\end{table}

\subsection{Spurious sources}

Given a detection probability threshold P the number of spurious
sources can be estimated by multiplying (1-P) for the number of
independent cells in the sky covered near a given flux limit.  Faint
fluxes ($F_X=5-7\times10^{-14}$ \cgs) can be probed only in the MECS
inner $<8$ arcmin, where the typical detection cell is of 0.004-0.005
deg$^2$. The skycoverage at these fluxes is of 1-3 degrees and
therefore the number of faint spurious sources should be $<0.6$.
Sources of flux of $3-10\times10^{-13}$ \cgs can be detected in the
whole MECS, with typical detection cells of 0.007-0.012 deg$^2$.  The
skycoverage at these fluxes is of 40 and 80 deg$^2$ and therefore, the
number of spurious sources should be between 2.5 and 3.5 In total the
fraction of spurious sources should be smaller than 5\%. These numbers
have been checked using simulations.

\subsection{Flux and sky coverage uncertainties}

The MECS absolute flux calibration has been obtained using several
observation of the Crab nebula in the three years of the BeppoSAX
mission.  Assuming a power-law model the energy index and the 2-10 keV
flux are found to be $\alpha_E = 1.088\pm 0.002 $ and
$F_{2-10keV}=2.008\pm 0.006 \times 10^{-8}$ erg cm$^{-2}$s$^{-1}$.
Repeated observations have not revealed any significant variations in
these parameters so far (Sacco 1999).  Main flux and sky coverage
uncertainties are due to the unknown spectrum of the sources near the
detection limit. We evaluated this uncertainty by calculating fluxes
and sky coverage in the two limiting caseses of power law spectra with
$\alpha_E=0.2$ and 1.0, see figure \ref{sky}b, in addition to the
default case (power law spectrum with $\alpha_E=0.6$).

\subsection{Source confusion}

Source confusion is likely to affect surveys performed with
instruments with limited spatial resolution like the BeppoSAX MECS.
It is therefore important to quantify the effect of source confusion
for the present survey. At 5-10 keV fluxes of $2.5-5-10\times10^{-14}$
\cgs we expect less than 60, 20 and 7 sources per square degree,
respectively, based on ASCA and BeppoSAX 2-10 keV surveys
(e.g. Cagnoni et al. 1998, Ueda et al.  1998, Giommi et al. 2000).
The probability to find two sources with comparable fluxes equal to
the above values within 2-4 arcmin (twice the size of the typical
slide-cells, see section 2.1) is $\sim0.4$\%, $\sim0.05$\% and
$\sim3$\% respectively. The fraction of confused sources with flux
$\ls1-2\times 10^{-13}$ \cgs, should therefore be smaller than 10\%
. A comparison with the ROSAT PSPC may be instructive.  The on-axis
PSPC PSF is a factor of $\approx 3$ sharper than the MECS one, but the
source density at the PSPC flux limit for a 50 ks observation is 10-20
times higher than that at the HELLAS flux limit (see e.g. Zamorani et
al. 1999), giving rise to similar confusion probabilities.
Simulations confirmed the reliability of the above numbers.

We have also looked at the spatial extension and asymmetry in the
counts distribution of the 147 sources to understand if any of these
sources can be actually due to the contribution of 2 or more confused
sources.  We find that this might be the case for about 20 sources
distributed in the whole range of fluxes. This is an upper limit to
the number of confused sources, since some of these may of course be
truly extended sources. In fact 6 of these sources have been already
identified with clusters of galaxies.  We will discuss this subsample
more in detail in a follow-up publication.  We conclude that confusion
is likely to alter by little the results on the logN-logS presented in
the next section.

\subsection {The integral logN-logS}

\begin{figure}[h]
\centerline{
\psfig{file=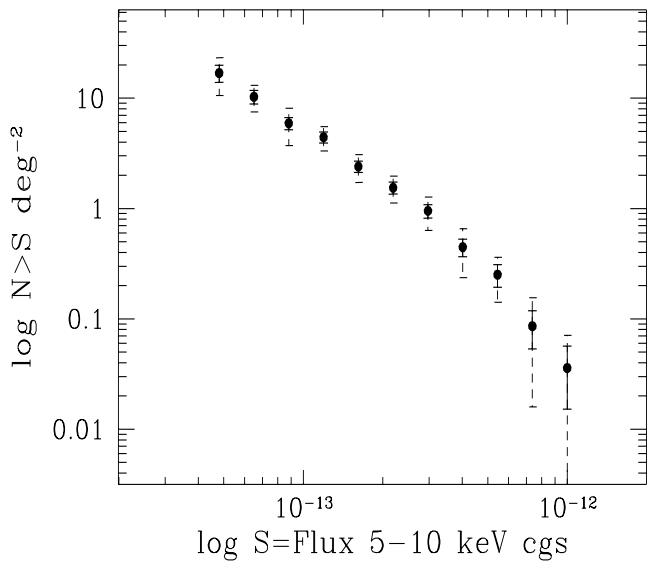, width=9cm, height=10cm, angle=-90}
}
\caption{ HELLAS 5-10 keV integral logN-logS.  Solid error bars = 1
$\sigma$ statistical confidence inteval.  Dashed error bars include
systematic uncertainties, see text.  
}
\label{lnls}
\end{figure}

Figure \ref{lnls} and table 1 present the integral 5-10 keV logN-logS
of the 147 HELLAS sources. First quoted errors in Table 1 and solid
error bars in figure \ref{lnls} are the $1\sigma$ statistical
confidence interval. Errors in brackets and dashed error bars in
figure include systematic uncertainties, mainly due to the lack of
knowledge of the real spectrum of the faint sources, see previous
section. We find 16.9$\pm$3.0(6.4) sources deg$^{-2}$ at $F_{5-10
keV}=4.8\times10^{-14}$ \cgs.  This number corresponds to a resolved
fraction of the 5-10 XRB equal to 20-30 \%, depending on the XRB
normalization (see Comastri 2000 and Vecchi et al. 1999).

\section{X--ray spectral analysis}

\begin{figure*}
\centerline{
\psfig{file=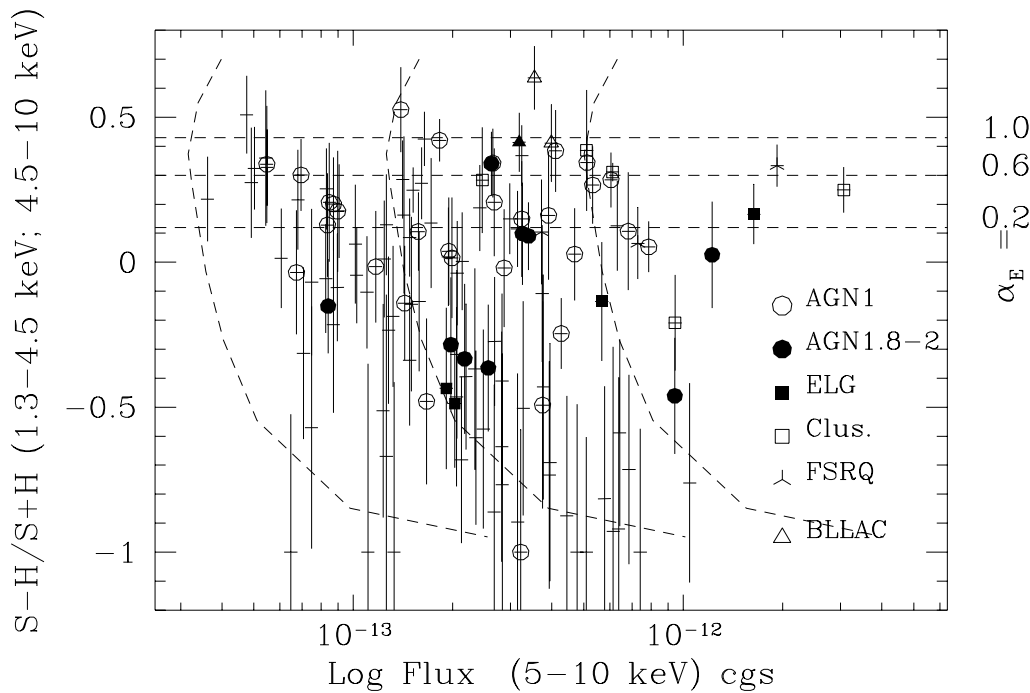, width=11cm, height=15cm, angle=-90}
}
\caption{
The softness ratio (S-H)/(S+H) plotted as a function of the 
5-10 keV flux for the 128 sources in the HELLAS sample not partially
or totally covered by the strongback support of the MECS window
(57 optically identified sources). Different symbols mark identified sources:
open circles = broad line quasars and Sy1; 
filled circles= type 1.8-1.9-2.0
AGN; filled squares= starburst galaxies and LINERS; open triangles=
BL Lacertae objects; skeletal triangles= Broad Line Radio-loud quasars; 
filled triangles= radio galaxies; open squares= clusters of galaxies.
Dashed lines mark loci of equal 4.5-10 keV count rate.
}
\label{hrtfx}
\end{figure*}

For many of the HELLAS sources the total number of counts detected is
$<$100, preventing the use of proper spectral fitting procedures to
study their spectrum.  The broad band X-ray spectral properties of the
HELLAS sources can however be investigated using count ratios.  We
have calculated for each source the softness ratio (L+M-H)/(L+M+H), or
(S-H)/(S+H) (L=1.3-2.5 keV band, M=2.5-4.5 keV band, S=L+M=1.3-4.5 keV
band, H=4.5-10 keV band), and the hardness ratios HR1=(M-L)/(M+L) and
HR2=(H-M)/(H+M).  MECS1 had a lower sensitivity at low energy (E$<$4
keV) than MECS 2 and 3, because of a thicker Kapton filter. Therefore
count ratios of sources observed with 3 MECS units have been corrected
for this effect. The correction is however small, always smaller than
the statistical errors.  Sources under or close to the berillium
strongback supporting the MECS window have been excluded from this
analysis, because their observed hardness may be systematically higher
than real. The number of the remaining sources in the sample used for
the following analysis is of 128.  (S-H)/(S+H) for these sources is
given in Table 2.
Figure \ref{hrtfx} plots (S-H)/(S+H) as a function of the source 5-10
keV flux. Many of the HELLAS sources have a low (S-H)/(S+H),
indicating a hard spectrum. Errors are however quite large. To
evaluate the number of sources with a softness ratio inconsistent with
that expected by an unobscured power law at a given confidence level,
we have compared the counts observed in the 1.3-4.5 keV band with that
predicted by a power law model, based on the 4.5-10 keV count
rates. Assuming $\alpha_E=0.6$, we find that 36 of the 128 sources
have 1.3-4.5 keV count rates lower than that expected at confidence
level $\gs 95 \%$.  Five sources have low band count rates lower than
expected at a confidence level $\gs 99.7 \%$. Large absorbing columns
densities are likely responsible for the hard spectrum of these
sources.  

Most of these very hard sources have flux higher than
$1-2\times10^{-13}$ \cgs. Indeed, a deficit of very hard sources at
lower fluxes is evident in figure \ref{hrtfx}. There is a possible
astrophysical reason for this deficit, namely a redshift effect: the
observed softness ratio of sources with similar intrinsic absorbing
column density increases with the redshift, as the observed cut-off energy
moves toward lower energies. On the other hand, a deficit of faint
hard sources can also be the result of a reduced sensitivity to hard
sources. This is due to the rapid increase of the vignetting of the
telescopes with the energy and with the off-axis angle. In other
words, the sky coverage decreases faster for the hardest sources than
for the softest ones. To quantify this effect we have computed loci
of equal 4.5-10 keV count rate for a given flux, shown by dashed lines
in figure \ref{hrtfx}. The strong curvature of these lines toward low
values of (S-H)/(S+H) indicate that most of the deficit of faint hard
sources is probably due to this effect.  The curves are bent toward
high flux values at high (S-H)/(S+H) too, because the MECS sensitivity
is reduced for very soft sources by the berillium window, which
absorbs most photons below $\sim2$ keV.  The 4.5-10 keV sensitivity is
maximum for an unabsorbed power law spectrum of $\alpha_E=0.6-0.8$.

Different symbols in figure \ref{hrtfx} mark optically identified
sources (Fiore et al. 1999, 2001, La Franca et 2001, in preparation).
Note as several of the narrow line AGN and emission line galaxies have
an hard X-ray, possibly absorbed, spectrum.  Intriquingly, also some
of the broad line AGN have an hard X-ray spectrum suggesting some
absorption also in these sources. See Fiore et al. 2001 and Comastri
et al. 2001 for a more detailed discussion.

\begin{figure}
\centerline{
\psfig{file=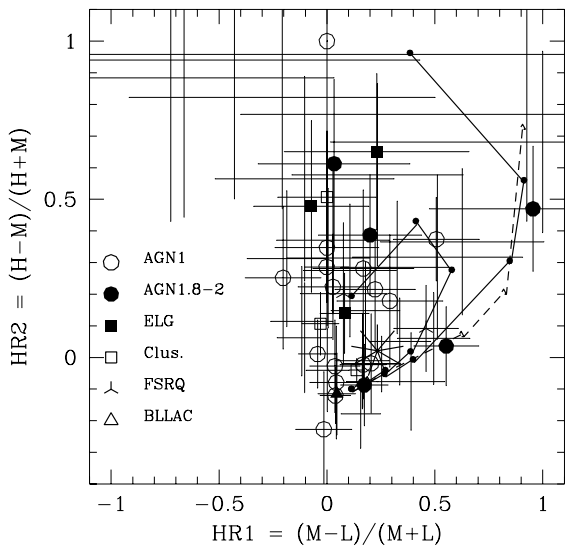, width=9cm, height=9cm}
}
\caption{ The hardness ratio HR1=(M-L)/(M+L) as a function of the hardness
ratio HR2=(H-M)/(H+M) for the 56 sources  not partially
or totally covered by the strongback support of the MECS window and
with signal to noise $>3.8$ (32 optically identified sources).  
Symbols as in figure \ref{hrtfx}.  The big star
represent the predicted position for an unabsorbed power law source
with $\alpha_E=0.4$. The rightmost solid curve 
represents the expectation of a power law model with $\alpha_E=0.8$
absorbed by a column of log$N_H$=0, 22, 22.7, 23, 23.7 and 24 (down to
up) at z=0.  The dashed curve represents the expectation of the same
models with the absorber at z=0.4.  The  innermost solid curve
represents the expectation of a partial covering model with Log$N_H$ as
in the previous cases and  covering fraction of 90\%.
}
\label{hrsr}
\end{figure}

The source spectra are likely to be more complex than a simple
absorbed power law. To study in more detail this complexity we have
computed the hardness ratios HR1 and HR2, plotted one against the
other in figure \ref{hrsr}. In this plot the analysis is limited to
the 56 source with signal to noise $>3.8$ (31 identified), not covered
by the MECS strongback.  Most of the points lie to the left of dashed
and first solid line in figure, (absorbed power law models),
suggesting that while substantial columns (log$N_H$=22-24) are likely
in the HELLAS sources, soft components spilling out below 3-5 keV are
also common in the sample, making softer HR2. A partial covering model
has been used to parameterize an absorbed spectrum with a soft
component.  The innermost curve represents the expectation of such a
model, with covering fraction of 90\%.  These models encompass most of
the points, taking into account the rather large errors on the
hardness ratios.  The extreme position of the few points in the upper
left part of the diagram is due to very few counts measured in the
middle band, which makes extremely high HR2 and extremely low HR1.
All these points are in any case consistent with the above mentioned
simple models, within the errors.  A more detailed analysis of the
soft X-ray properties of the HELLAS sources is presented by Vignali et
al. (2001) where the study of the X-ray spectrum is extended down to
0.5 keV using PSPC and HRI detections or opper limits.

\section{Discussion and Conclusions}

About 84 deg$^2$ of sky have been surveyed in the hard 5-10 keV band
using 142 BeppoSAX MECS independent pointings.  A statistically well
defined and flux limited sample of 147 hard X-ray selected sources was
assembled, and used to estimate the 5--10 keV logN--logS.  The number
counts are affected by both statistical and systematic errors.  The
latter, mainly due to the lack of knowledge on the spectrum of faint
sources, have been estimated assuming a range of spectral slopes in
the count rate to flux conversion.  The source surface density ($\sim$
16.9 $\pm$ 6.4 deg$^{-2}$) at the survey flux limit of
$4.8\times10^{-14}$ \cgs corresponds to a resolved fraction of the
5--10 keV XRB of the order of 20--30 \% depending on the XRB
normalization (Vecchi et al. 1999, Comastri 2000).

Hardness ratios have been used to study the X-ray spectrum of the
HELLAS sources. These hardness ratios indicate rather hard spectra,
harder than in previous 0.7-10 keV ASCA surveys (Ueda et al. 1999,
Della Ceca et al. 1999).  This hardness may be due to substantial
absorbing columns.  In fact, a large fraction of identified type 1.8-2
AGN show softness ratios similar to those expected from power law
models reduced at low energy by column densities of log$N_H$=22--24,
at the source redshift (see e.g. Fiore et al. 2000, 2001 and Comastri
et al. 2001).

Many of the sources may have a softer component emerging below 3--4
keV. Indeed the cross--correlation between the HELLAS catalogue and
the ROSAT PSPC public archive data (Vignali et al. 2001) indicates
that about two third of the HELLAS sources in the field of view of a
ROSAT observation have a counterpart in the soft (0.5--2 keV) energy
range The nature of such a component is still not clear and could be
due to nuclear photons spilling from a partial covering screen or
reflected from a warm-hot medium, or to extended extranuclear
components, possibly related to circumnuclear starburst regions and
winds. The presence of a soft component not directly related to the
nuclear emission might easily lead to a wrong estimate of the
intrinsic soft X-ray luminosity, especially for heavily absorbed
sources.  According with the XRB baseline model, absorbed AGN become
progressively more important towards faint fluxes and thus spurious
evolutionary terms can be introduced in the luminosity function
derived from soft X--ray selected samples.  An estimate of the
importance of such a bias is not straighforward and also strongly
model--dependent (see Miyaji et al. 2000 for a detailed
discussion). The optical identification of a sizeable sample of hard
X--ray selected sources coupled with more sensitive soft and hard
X--ray surveys like those that will be performed in the near future by
{\it Chandra} and XMM--{\it Newton} will provide new insights in this
direction.

\section{Appendix: Position Accuracy}

The location of an X-ray source by the BeppoSAX MECS is affected
by statistical and systematical uncertainties. 

The minimum number of counts in an HELLAS source is about 30, against
about 100 backgrounds counts.  For such faint sources, given the MECS
PSF, the statistical uncertainty on the centroid is of 30-40 arcsec.
For bright sources the uncertainty on the source centroid due to the
PSF is about 13 arcsec ($1\sigma$), measured using 12 observations of
LMC X-3.  On the other hands, systematic errors can often be larger
the these figures.  The main sources of systematic errors in BeppoSAX
positions are:

\begin{enumerate}

\item
The complete unavailability of the star-tracker ``z'' (the one
co-aligned with the X-ray telescopes) in about 10 \% of the
observations. In these cases the error on the pointing position
reconstruction can be as big as 2 arcmin.

\item
The unavailability of the star-tracker ``z'' for part of each orbit in
most observations.  This often produces ``jumps'' in the attitude
reconstruction when passing control from one star-tracker to another.
These ``jumps'' are due to a non perfect calibration of the
misalignement between the three star-trackers. An error of about 20
arcsec in the on-board calibration of the misallignement has been
discovered during summer 1997. It has been corrected in the on-board
software at the beginning of May 1999. Observations performed after
May 5th 1999 should provide more accurate positions than earlier
observations.

\item
Since August 1997 BeppoSAX is operating in the so called ``1 gyro
mode''. For part of the orbit the satellite attitude is controlled by
one star-tracker and one gyro. All configuarations involving
star-trackers ``y'' and ``-x'' plus the gyro produce a pointing
accuracy worse than the configurations including the star-trackers
``y'' and ``-x'' simultaneously in the control loop, and those (the
best) in which the star-tracker ``z'' is in the control loop.

\end{enumerate}

\begin{figure}
\centerline{
\psfig{file=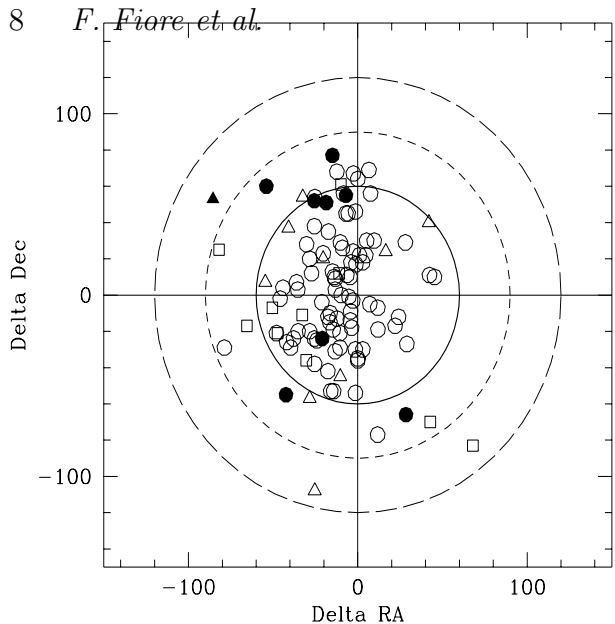, width=8cm, height=8cm}
}
\caption{ The deviation in RA and dec between the MECS position and
the catalog position of 107 AGN (targets or serendipitously found in
the fields) in the HELLAS survey.  Filled symbols: sources observed
without ``z'' startracker in the control loop. Open symbols: sources
observed with ``z'' startracker for at least part of the time.
Circles: sources detected at off-axis angles $<3$ arcmin.  Squares:
sources detected at off-axis angles between 7 and 15 arcmin. Triangles
source detected outside 15 arcmin.  The thick solid circle has a
radius of 1 arcmin, the short-dashed and long-dashed circles have
radii of 1.5 and 2 arcmin respectively.  
}
\label{pos}
\end{figure}

Figure \ref{pos} shows the deviation in RA and Dec between the MECS
sky positions and the optical (or radio) positions of a sample of 107
known AGN (targets or serendipitously found in the HELLAS fields).
While the median deviation on dec is zero, MECS RA are systematically
smaller than catalog ones by $\Delta RA\sim15$ arcses.  The figure
illustrates that when star-tracker ``z'' is not in the control loop
the typical error on the position is between 1 and 1.5 arcmin.
Sources detected at large off-axis angles show larger deviations
because of the degradation of the MEVS PSF: the radii encompassing 67
\% and 90 \% of the 79 sources observed with z startracker in use and
detected at off axis angle $<3$ arcmin are 43 arcsec and 56 arcsec
respectively.  The same radii for the 19 sources detected at off-axis
angles in the range 7-23 are 63 arcsec and 85 arcsec respectively.

Source positions in Table 2 were corrected for the above systematic effects
by performing a simple boresight correction with respect to the
known positions of the target and of any other known bright source
in each field.

\bigskip
\centerline{\bf Acknowledgements}

We thank the BeppoSAX SDC, SOC and OCC teams for the successful
operation of the satellite and preliminary data reduction and
screaning, A. Matteuzzi for his work on MECS source position
reconstruction, D. Ricci and M. Capalbi for the help with the BeppoSAX
archive and databases. We also thank R. Maiolino, F. Pompilio,
M. Perri, M. Salvati and G. Zamorani for useful discussions.
This research has been partially supported by ASI contract
ARS--99--75, MURST grants Cofin--98--032, Cofin--99--034, Cofin--00-02-36,
and a 1999 CNAA grant.


\begin{table*}
\caption{\bf HELLAS sample$^a$}
\begin{tabular}{lcccccccc}
\hline
RA (2000) & Dec (2000) & offset & Flux 5-10keV & SNR  & S-H/S+H & HR1 & HR2 & ID  \\
          &            & arcmin & $10^{-13}$\cgs &     &         &     &     &  \\
\hline
0 26 36.5 & -19 44  13 &   13.2 &  3.39 &  5.5 &  0.09 $\pm$0.12  &  0.55$\pm$0.15 &  0.04$\pm$0.12 & 120 \\ 
0 27  9.9 & -19 26  31 &    6.2 &  1.83 &  6.5 &  0.42 $\pm$0.07  &  0.04$\pm$0.07 & -0.12$\pm$0.09 & 124 \\ 
0 27 43.9 & -19 30  29 &   11.0 &  1.58 &  3.4 &       -b-       &  -c- &  -c- & 121 \\ 
0 45 49.6 & -25 15  13 &   23.7 &  3.26 &  3.3 &  0.10 $\pm$0.18  &  -c- &  -c- & 103\_1 \\ 
0 48  5.8 & -25 4  32 &   14.9 &  1.48 &  3.6 & -0.34  $\pm$0.23  &  -c- &  -c- & 107 \\ 
1 18  3.9 & 89 20  12 &    8.3 &  0.87 &  3.3 & -0.21  $\pm$0.30  &  -c- &  -c- & 431 \\ 
1 21 56.8 & -58 44   5 &   15.6 &  2.57 &  4.8 & -0.36 $\pm$0.22  &  0.03$\pm$0.35 &  0.61$\pm$0.27 & 39 \\ 
1 34 14.3 & -29 45  41 &   14.2 &  1.50 &  3.3 & -0.14 $\pm$0.30  &  -c- &  -c- & 99 \\ 
1 34 28.6 & -30 6  34 &   12.2 &  1.33 &  3.5 &         -b-      &  -c- &  -c- & 90 \\ 
1 34 33.3 & -29 58  38 &    6.0 &  0.87 &  3.9 &  0.20 $\pm$0.16  & -0.20$\pm$0.18 &  0.25$\pm$0.23 & 93 \\ 
1 34 49.6 & -30 2  34 &    5.4 &  0.71 &  3.5 & -0.31  $\pm$0.30  &  -c- &  -c- & 91 \\ 
1 35 30.2 & -29 51  21 &    8.4 &  0.90 &  3.2 &  0.18 $\pm$0.21  &  -c- &  -c- & 96 \\ 
1 40  8.9 & -67 48  13 &    8.0 &  2.83 &  3.2 & -0.41 $\pm$0.30  &  -c- &  -c- & 20 \\ 
1 53  3.9 & 89 12  20 &    2.4 &  0.54 &  3.1 &  0.36  $\pm$0.23  &  -c- &  -c- & 426 \\ 
2 42  1.8 & 0 0  46 &   10.0 &  1.47 &  3.7 &          -b-       &  -c- &  -c- & 172 \\ 
2 42  9.4 & 0 2  29 &    7.5 &  0.68 &  3.2 &  0.22    $\pm$0.17  &  -c- &  -c- & 174 \\ 
3 8  8.5 & 89 8  41 &    4.9 &  0.65 &  3.5 & -1.00    $\pm$0.48  &  -c- &  -c- & 424 \\ 
3 8 19.0 & 2 46  27 &   18.5 &  5.09 &  3.0 & -1.00    $\pm$0.40  &  -c- &  -c- & 195 \\ 
3 15 45.0 & -55 29  26 &   14.5 &  2.65 &  4.6 &  0.34 $\pm$0.12  &  0.04$\pm$0.12 & -0.03$\pm$0.15 & 45 \\ 
3 17 32.4 & -55 20  12 &   21.0 &  4.10 &  3.8 &  0.38 $\pm$0.14  &  -c- &  -c- & 46 \\ 
3 33  9.6 & -36 19  40 &   15.2 &  3.98 &  3.3 &  0.41 $\pm$0.13  &  -c- &  -c- & 72 \\ 
3 34  7.4 & -36 4  22 &    5.1 &  1.95 &  3.9 &  0.04  $\pm$0.19  &  0.29$\pm$0.24 &  0.18$\pm$0.21 & 75 \\ 
3 36 51.3 & -36 15  57 &   14.7 &  3.72 &  3.9 &  0.10 $\pm$0.18  &  0.08$\pm$0.20 &  0.20$\pm$0.22 & 73 \\ 
4 32 27.9 & 5 13   5 &   15.0 &  2.06 &  3.3 & -0.32   $\pm$0.29  &  -c- &  -c- & 203 \\ 
4 37 14.5 & -47 30  58 &   16.0 &  2.68 &  3.5 &  0.21 $\pm$0.19  &  -c- &  -c- & 53 \\ 
4 38 47.9 & -47 29   6 &   20.1 &  4.69 &  4.6 &  0.03 $\pm$0.16  &  0.22$\pm$0.19 &  0.22$\pm$0.19 & 54 \\ 
5 2 15.5 & 12 4   7 &   20.9 &  7.41 &  3.8 & -1.00    $\pm$0.43  &  -c- &  -c- & 220 \\ 
5 15 13.7 & 1 8   7 &    7.6 &  1.26 &  3.3 &  0.13    $\pm$0.17  &  -c- &  -c- & 185 \\ 
5 20 48.3 & -45 42   0 &    9.6 &  5.90 &  4.0 &         -b-     &-b-            &    -b-        & 57 \\ 
5 48 41.3 & -60 52  18 &   19.5 &  2.42 &  3.4 &  0.19 $\pm$0.15  &  -c- &  -c- & 32 \\ 
5 50  0.2 & -61 2  22 &    5.9 &  0.90 &  4.2 & -0.09  $\pm$0.19  &  0.33$\pm$0.26 &  0.28$\pm$0.21 & 26 \\ 
5 52  6.1 & -60 59  48 &   11.1 &  1.23 &  3.3 &        -b-      &  -c- &  -c- & 30 \\ 
5 52 51.3 & -60 57  18 &   17.1 &  2.06 &  3.9 & -0.47 $\pm$0.31  & -0.66$\pm$0.70 &  0.88$\pm$0.44 & 31 \\ 
6 23 56.6 & -69 21  13 &    6.1 &  1.33 &  3.0 & -1.00 $\pm$0.59  &  -c- &  -c- & 6 \\ 
6 25 31.3 & -69 19   9 &    6.4 &  1.42 &  3.0 &  0.16 $\pm$0.27  &  -c- &  -c- & 8 \\ 
6 46 39.3 & -44 15  35 &   16.6 &  4.27 &  6.9 & -0.25 $\pm$0.12  &  0.51$\pm$0.20 &  0.37$\pm$0.13 & 65 \\ 
6 46 42.7 & -44 32  29 &   19.0 &  2.68 &  4.0 & -0.27 $\pm$0.22  &  0.63$\pm$0.38 &  0.37$\pm$0.23 & 58 \\ 
6 55 39.6 & 79 10  48 &    4.6 &  0.75 &  2.9 & -0.57  $\pm$0.42  &  -c- &  -c- & 405 \\ 
7 21 29.6 & 71 14   4 &    8.1 &  0.84 &  3.5 &  0.13  $\pm$0.18  &  -c- &  -c- & 385 \\ 
7 41 40.3 & 74 14  57 &   22.6 & 30.66 &  7.5 &  0.25  $\pm$0.08  & -0.03$\pm$0.09 &  0.11$\pm$0.10 & 392 \\ 
7 41 45.2 & 74 26  23 &   13.1 &  3.74 &  3.2 & -0.11  $\pm$0.24  &  -c- &  -c- & 393 \\ 
7 43  9.1 & 74 29  19 &    7.2 &  6.05 &  6.3 &  0.28  $\pm$0.09  &  0.17$\pm$0.11 & -0.02$\pm$0.11 & 394 \\ 
8 37 37.2 & 25 47  48 &   12.1 &  2.63 &  3.9 &      -b-         & -b-           &   -b-         & 243 \\ 
8 38 59.9 & 26 8  13 &   23.0 & 16.38 &  6.0 &  0.17   $\pm$0.10  &  0.08$\pm$0.11 &  0.14$\pm$0.13 & 246 \\ 
9 46  5.3 & -14 2  59 &   15.8 &  2.82 &  3.8 & -0.64  $\pm$0.40  &  -c- &  -c- & 138 \\ 
9 46 17.9 & -14 10  27 &   10.9 &  2.04 &  3.5 &     -b-         &  -c- &  -c- & 136 \\ 
9 46 32.8 & -14 6  15 &   16.4 &  3.22 &  4.1 & -1.00  $\pm$0.42  &  -c- &  1.00$\pm$0.42 & 137 \\ 
10 29 19.1 & 50 48  15 &   17.5 &  5.78 &  3.0 & -0.82 $\pm$0.39  &  -c- &  -c- & 299 \\ 
10 32 15.8 & 50 51   3 &   10.4 &  3.13 &  2.6 &     -b-         &  -c- &  -c- & 300 \\ 
10 34 43.1 & 39 29  18 &    8.4 &  1.11 &  3.1 & -1.00 $\pm$0.65  &  -c- &  -c- & 273 \\ 
10 34 52.0 & 39 40  12 &    5.3 &  1.41 &  4.4 &  0.29 $\pm$0.13  & -0.12$\pm$0.14 &  0.11$\pm$0.17 & 278 \\ 
10 52 45.4 & 57 30  42 &    4.0 &  0.83 &  4.4 & -0.06 $\pm$0.19  &  0.03$\pm$0.26 &  0.37$\pm$0.23 & 321 \\ 
10 54 19.8 & 57 25   9 &   13.4 &  2.62 &  5.0 &  0.34 $\pm$0.11  &  0.17$\pm$0.11 & -0.09$\pm$0.13 & 319 \\ 
10 54 21.7 & 57 36  24 &   16.6 &  2.13 &  4.0 & -0.68 $\pm$0.29  &  1.00$\pm$0.98 &  0.68$\pm$0.29 & 323 \\ 
\hline

\end{tabular}

 
\end{table*}

\setcounter{table}{1}

\begin{table*}
\caption{\bf HELLAS sample$^a$ }
\begin{tabular}{lcccccccc}
\hline
RA (2000) & Dec (2000) & offset & Flux 5-10keV & SNR  & S-H/S+H & HR1 & HR2 & ID  \\
          &            & arcmin & $10^{-13}$\cgs &     &         &     &     &  \\
\hline
11 1 46.4 & 72 26  11 &   22.3 &  7.29 &  4.9 &  0.07  $\pm$0.12  &  0.46$\pm$0.15 &  0.09$\pm$0.14 & 387 \\ 
11 2 37.2 & 72 46  38 &   20.7 &  7.87 &  7.7 &  0.05  $\pm$0.09  & -0.00$\pm$0.10 &  0.29$\pm$0.11 & 390 \\ 
11 6 14.0 & 72 43  16 &    8.5 &  1.67 &  3.6 & -0.48  $\pm$0.29  &  -c- &  -c- & 389 \\ 
11 7  4.9 & -18 16  28 &    7.7 &  1.23 &  3.1 & -0.51 $\pm$0.37  &  -c- &  -c- & 131 \\ 
11 18 11.9 & 40 28  33 &    4.2 &  0.85 &  2.9 &  0.21 $\pm$0.20  &  -c- &  -c- & 283 \\ 
11 18 46.2 & 40 27  39 &    4.8 &  1.39 &  3.9 &  0.53 $\pm$0.15  & -0.01$\pm$0.13 & -0.23$\pm$0.18 & 282 \\ 
11 34 52.7 & 70 23   9 &   15.5 &  3.77 &  3.1 & -0.43 $\pm$0.39  &  -c- &  -c- & 380 \\ 
11 56 39.2 & 65 17  57 &    5.7 &  0.75 &  3.2 & -0.07 $\pm$0.25  &  -c- &  -c- & 372 \\ 
11 57  1.7 & 65 27  24 &   15.4 &  2.14 &  3.7 &  0.00 $\pm$0.17  &  -c- &  -c- & 374 \\ 
12 4  7.6 & 28 8  30 &   16.1 &  5.10 &  3.0 &  0.39   $\pm$0.21  &  -c- &  -c- & 252 \\ 
12 17 45.1 & 47 29  55 &   16.4 &  3.24 &  5.0 &  0.37 $\pm$0.10  &  0.00$\pm$0.10 & -0.04$\pm$0.13 & 292 \\ 
12 17 50.3 & 30 7   8 &   19.8 &  3.54 &  3.2 &  0.64  $\pm$0.11  &  -c- &  -c- & 265 \\ 
12 18 55.0 & 29 58  12 &   12.7 &  1.98 &  3.3 & -0.28 $\pm$0.23  &  -c- &  -c- & 264 \\ 
12 19 21.6 & 47 11   7 &    9.4 &  1.32 &  3.2 &      -b-        &  -c- &  -c- & 288 \\ 
12 19 45.7 & 47 20  42 &    7.5 &  1.17 &  4.0 & -0.02 $\pm$0.19  &  0.00$\pm$0.24 &  0.35$\pm$0.25 & 290 \\ 
12 22  6.8 & 75 26  17 &    6.5 &  2.46 &  3.5 &  0.28 $\pm$0.18  &  -c- &  -c- & 400 \\ 
12 29 23.7 & 1 51  38 &   14.1 &  1.61 &  4.3 &  0.27  $\pm$0.12  &  0.17$\pm$0.12 & -0.01$\pm$0.15 & 186 \\ 
12 40 26.0 & -5 13  20 &   11.7 &  3.13 &  4.3 &     -b-         &  0.17$\pm$0.16 &  0.10$\pm$0.17 & 167 \\ 
12 40 29.6 & -5 7  46 &   16.8 &  1.92 &  3.6 & -0.44  $\pm$0.28  &  -c- &  -c- & 169 \\ 
12 54 28.0 & 59 21   1 &   24.3 &  6.37 &  2.9 & -0.92 $\pm$0.52  &  -c- &  -c- & 324 \\ 
12 55 16.6 & -5 39  22 &   16.3 &  2.20 &  4.0 & -0.39 $\pm$0.25  &  0.23$\pm$0.40 &  0.58$\pm$0.29 & 161 \\ 
12 56  9.9 & -5 54  30 &    7.6 &  0.91 &  3.6 &  0.18 $\pm$0.16  &  -c- &  -c- & 157 \\ 
13 4 24.3 & -10 23  53 &    4.4 &  1.28 &  3.9 & -0.23 $\pm$0.25  & -0.10$\pm$0.41 &  0.56$\pm$0.32 & 151 \\ 
13 4 38.2 & -10 15  47 &    5.9 &  1.43 &  3.8 & -0.14 $\pm$0.20  &  -c- &  -c- & 155 \\ 
13 4 45.1 & -5 33  37 &    7.5 &  1.26 &  3.0 & -0.67  $\pm$0.56  &  -c- &  -c- & 162 \\ 
13 5 32.3 & -10 32  35 &   22.0 & 19.27 &  7.2 &  0.33 $\pm$0.07  &  0.16$\pm$0.07 & -0.08$\pm$0.09 & 150 \\ 
13 5 36.5 & -5 43  30 &   22.7 &  6.42 &  3.4 & -0.59  $\pm$0.32  &  -c- &  -c- & 160 \\ 
13 36 34.3 & -33 57  47 &   21.8 &  3.18 &  4.6 &  0.41$\pm$0.10  &  0.05$\pm$0.09 & -0.12$\pm$0.13 & 84 \\ 
13 38 34.1 & 48 21   5 &    4.3 &  1.24 &  3.3 &  0.01 $\pm$0.20  &  -c- &  -c- & 295 \\ 
13 42 47.9 & 0 21   9 &   16.9 &  2.48 &  3.4 & -0.58  $\pm$0.34  &  -c- &  -c- & 180 \\ 
13 42 59.3 & 0 1  38 &   20.6 &  3.25 &  3.2 &  0.15   $\pm$0.20  &  -c- &  -c- & 176 \\ 
13 48 20.8 & -30 11   6 &   14.4 &  2.18 &  3.5 & -0.33$\pm$0.26  &  -c- &  -c- & 100 \\ 
13 48 24.3 & -30 25  47 &   14.8 &  3.15 &  4.7 &  0.11$\pm$0.13  &  0.18$\pm$0.16 &  0.15$\pm$0.15 & 94 \\ 
13 48 37.9 & -30 9  11 &   12.3 &  1.59 &  3.2 &     -b-         &  -c- &  -c- & 101 \\ 
13 48 45.4 & -30 29  36 &   14.4 &  5.11 &  7.1 &  0.34$\pm$0.09  & -0.04$\pm$0.09 &  0.01$\pm$0.11 & 92 \\ 
13 50  9.4 & -30 19  55 &   10.8 &  5.08 &  7.4 &    -b-         &     -b-       &   -b-         & 97 \\ 
13 53 54.6 & 18 20  33 &   17.5 &  6.82 &  3.4 &  0.11 $\pm$0.20  &  -c- &  -c- & 228 \\ 
13 55 54.1 & 18 13  35 &   19.8 &  6.14 &  2.8 & -0.93 $\pm$0.64  &  -c- &  -c- & 226 \\ 
14 11 58.7 & -3 7   2 &   20.7 &  3.93 &  3.6 & -0.73  $\pm$0.39  &  -c- &  -c- & 171 \\ 
14 17 12.5 & 24 59  28 &   12.9 &  0.69 &  3.7 &  0.30 $\pm$0.12  &  -c- &  -c- & 239 \\ 
14 18 31.1 & 25 11   7 &    8.6 &  6.11 & 18.7 &  0.31 $\pm$0.03  &  0.14$\pm$0.03 & -0.04$\pm$0.04 & 241 \\ 
14 38 30.1 & 64 30  25 &   11.2 &  2.57 &  4.2 &   -b-           &  -b-          &-b-            & 364 \\ 
14 48 21.8 & -69 20  30 &   11.7 &  4.25 &  5.0 &  -b-           &   -b-         & -b-           & 10 \\ 
15 19 39.9 & 65 35  46 &   13.9 &  9.43 &  5.2 & -0.46 $\pm$0.20  &  0.95$\pm$0.48 &  0.47$\pm$0.20 & 375 \\ 
15 28 46.0 & 19 45  10 &    4.7 &  1.65 &  5.2 &  0.43 $\pm$0.09  &  0.16$\pm$0.09 & -0.18$\pm$0.11 & 230\_2 \\ 
15 28 47.3 & 19 39  10 &    5.0 &  1.57 &  5.3 &  0.11 $\pm$0.13  &  0.03$\pm$0.16 &  0.22$\pm$0.17 & 230\_1 \\ 
16 26 56.8 & 55 13  24 &   14.2 &  3.15 &  3.1 & -0.90 $\pm$0.51  &  -c- &  -c- & 305 \\ 
16 26 59.9 & 55 28  20 &   10.5 & 12.09 &  8.3 &                  &                &                & 307 \\ 
16 34 10.7 & 59 37  44 &    6.9 &  0.50 &  3.3 &  0.32 $\pm$0.14  &  -c- &  -c- & 325 \\ 
16 34 11.0 & 59 48  15 &    5.2 &  0.48 &  3.5 &  0.51 $\pm$0.13  &  -c- &  -c- & 328 \\ 
16 34 11.8 & 59 45  29 &    3.4 &  0.84 &  5.9 & -0.15 $\pm$0.16  &  0.20$\pm$0.24 &  0.39$\pm$0.19 & 327 \\ 
16 49 57.9 & 4 53  32 &   19.8 &  9.45 &  5.0 & -0.21  $\pm$0.16  &  0.00$\pm$0.23 &  0.51$\pm$0.21 & 201 \\ 
16 50 40.1 & 4 37  17 &   24.5 & 12.25 &  3.3 &  0.03  $\pm$0.18  &  -c- &  -c- & 200 \\ 
16 52 12.5 & 2 11  29 &   15.8 &  2.36 &  4.2 & -0.60  $\pm$0.30  & -0.21$\pm$0.71 &  0.82$\pm$0.35 & 187 \\ 
16 52 38.0 & 2 22  18 &    4.6 &  0.67 &  3.5 & -0.04  $\pm$0.21  &  -c- &  -c- & 190 \\ 
16 54 41.1 & 40 2  10 &   17.6 &  6.32 &  4.1 &  0.13  $\pm$0.16  & -0.19$\pm$0.18 &  0.31$\pm$0.21 & 279 \\ 
\hline

\end{tabular}

 
\end{table*}

\setcounter{table}{1}

\begin{table*}
\caption{\bf HELLAS sample$^a$ }
\begin{tabular}{lcccccccc}
\hline
RA (2000) & Dec (2000) & offset & Flux 5-10keV & SNR  & S-H/S+H & HR1 & HR2 & ID  \\
          &            & arcmin & $10^{-13}$\cgs &     &         &     &     &  \\
\hline
17 40 10.7 & 67 42  50 &   17.7 &  3.28 &  3.0 & -0.50 $\pm$0.37  &  -c- &  -c- & 375\_1 \\ 
17 42 36.3 & 68 0  44 &   10.6 &  2.76 &  3.6 &    -b-           &  -c- &  -c- & 375\_2 \\ 
17 50 25.4 & 60 56   1 &   11.7 &  2.10 &  3.2 &  -b-            &  -c- &  -c- & 340\_2 \\ 
17 51 30.3 & 61 0  43 &    6.7 &  0.55 &  4.0 &  0.33  $\pm$0.13  & -0.10$\pm$0.13 &  0.06$\pm$0.17 & 346\_2 \\ 
17 52 38.3 & 61 5  47 &    3.2 &  0.36 &  3.3 &  0.22  $\pm$0.15  &  -c- &  -c- & 351\_1 \\ 
17 53 49.0 & 60 59  52 &   12.7 &  1.10 &  3.8 & -0.10 $\pm$0.19  &  -c- &  -c- & 346\_1 \\ 
18 3 51.8 & 61 10  21 &    2.9 &  0.60 &  3.8 &  0.01  $\pm$0.17  &  0.11$\pm$0.22 &  0.27$\pm$0.21 & 353 \\ 
18 15 17.5 & 49 44  51 &    9.8 &  3.48 &  3.8 &   -b-             &        -b-      &  -b-            & 296 \\ 
18 18 58.6 & 61 14  42 &   18.6 &  2.07 &  3.6 & -0.04 $\pm$0.18  &  -c- &  -c- & 354 \\ 
18 19 18.4 & 60 56   6 &    3.5 &  1.52 &  8.1 &  0.25 $\pm$0.08  &  0.25$\pm$0.09 & -0.02$\pm$0.09 & 341 \\ 
18 19 35.5 & 60 58  46 &    2.1 &  0.49 &  3.3 &  0.27 $\pm$0.19  &  -c- &  -c- & 345 \\ 
18 19 39.5 & 60 53  26 &    3.7 &  0.83 &  5.0 &  0.25 $\pm$0.14  &  0.39$\pm$0.16 & -0.08$\pm$0.15 & 337 \\ 
18 36 11.3 & -65 7  20 &   22.8 &  4.44 &  3.3 & -0.88 $\pm$0.41  &  -c- &  -c- & 25 \\ 
20 42 47.6 & -10 38  30 &   21.1 &  5.32 &  4.6 &  0.27$\pm$0.13  &  0.20$\pm$0.13 & -0.02$\pm$0.16 & 147 \\ 
20 44 34.8 & -10 27  34 &   15.3 &  1.99 &  3.3 &  0.02$\pm$0.21  &  -c- &  -c- & 149 \\ 
21 22 59.2 & 89 1  58 &   15.3 &  4.77 &  3.1 & -1.00  $\pm$0.51  &  -c- &  -c- & 413 \\ 
21 38  8.9 & -14 33  13 &    7.0 &  2.34 &  3.4 & -0.37$\pm$0.35  &  -c- &  -c- & 134 \\ 
21 42 46.7 & 89 35  30 &   23.5 & 10.47 &  4.2 & -0.76 $\pm$0.34  &  0.93$\pm$1.33 &  0.77$\pm$0.34 & 432 \\ 
21 59 52.1 & 88 54  53 &   17.8 &  6.85 &  3.4 & -0.72 $\pm$0.33  &  -c- &  -c- & 410 \\ 
22 3  0.5 & -32 4  18 &   16.5 &  2.83 &  3.3 & -0.77  $\pm$0.45  &  -c- &  -c- & 85 \\ 
22 26 30.3 & 21 11  56 &   13.7 &  3.91 &  2.9 &  0.16 $\pm$0.22  &  -c- &  -c- & 237 \\ 
22 31 49.6 & 11 32   8 &   15.2 &  1.94 &  4.2 &  0.02 $\pm$0.17  &  0.01$\pm$0.19 &  0.31$\pm$0.22 & 213 \\ 
22 41 22.1 & 29 42  41 &   15.6 &  2.98 &  5.0 &  0.15 $\pm$0.12  &  0.34$\pm$0.14 &  0.05$\pm$0.14 & 256 \\ 
22 42 51.5 & 29 35  32 &    7.6 &  1.02 &  3.6 & -0.05 $\pm$0.17  &  -c- &  -c- & 254 \\ 
22 44 11.4 & 29 51  14 &   23.1 &  3.95 &  3.5 & -0.69 $\pm$0.41  &  -c- &  -c- & 258 \\ 
23 2 30.1 & 8 37   6 &   17.6 &  2.67 &  3.9 & -0.86   $\pm$0.44  & -0.43$\pm$2.64 &  0.96$\pm$0.46 & 206 \\ 
23 2 36.2 & 8 56  42 &   10.1 &  3.17 &  5.2 &       -b-         &  -b-          &   -b-         & 212 \\ 
23 6 59.2 & 8 48  40 &    4.1 &  1.02 &  3.0 &  0.06   $\pm$0.20  &  -c- &  -c- & 210 \\ 
23 15 36.4 & -59 3  40 &    8.3 &  2.03 &  4.6 & -0.49 $\pm$0.22  &  0.23$\pm$0.43 &  0.65$\pm$0.25 & 37 \\ 
23 16  9.8 & -59 11  24 &    6.1 &  1.32 &  3.9 & -0.19$\pm$0.24  &  0.51$\pm$0.39 &  0.32$\pm$0.26 & 35 \\ 
23 19 22.1 & -42 41  50 &   21.4 &  5.67 &  3.8 & -0.14$\pm$0.20  & -0.07$\pm$0.27 &  0.48$\pm$0.27 & 66 \\ 
23 27 28.7 & 8 49  30 &    7.1 &  0.55 &  3.0 &  0.34  $\pm$0.20  &  -c- &  -c- & 209 \\ 
23 27 37.1 & 8 38  56 &    7.9 &  1.48 &  5.5 &  0.09  $\pm$0.13  &  0.49$\pm$0.17 &  0.06$\pm$0.15 & 207 \\ 
23 29  2.4 & 8 34  39 &   20.0 &  2.87 &  4.2 & -0.02  $\pm$0.20  &  0.17$\pm$0.23 &  0.28$\pm$0.25 & 205 \\ 
23 31 55.6 & 19 38  34 &   17.0 &  3.75 &  3.0 & -0.49 $\pm$0.36  &  -c- &  -c- & 229 \\ 
23 55 32.7 & 28 35  11 &    7.7 &  1.72 &  3.2 &  0.14 $\pm$0.22  &  -c- &  -c- & 249 \\ 
23 55 53.3 & 28 36   5 &   12.0 &  4.17 &  4.0 &    -b-           & -b-  &  -b- & 250  \\ 
\hline

\end{tabular}

$^a$ This table is also available at the following URL:
http://argos.mporzio.astro.it/hellas
 
$^b$ partly obscured by the MECS strongback; -c- SNR$<3.8$

\end{table*}

\end{document}